\documentstyle[psfig,epsfig,rotating]{basi}
\def\arcmin{\hbox{$^\prime$}}
\def\arcsec{\hbox{$^{\prime\prime}$}}
\begin{document}
\title[Optical Observations] 
{Optical Observations and Multiband Modelling of the Afterglow of GRB041006 : Evidence of A 
Hard Electron Energy Spectrum}\author[Misra et al.] 
       {Kuntal Misra$^{1}$\thanks{e-mail : kuntal@aries.ernet.in}
, L. Resmi$^{2,3}$, S. B. Pandey$^{1}$, D. Bhattacharya$^{2}$ and R. Sagar$^{1}$\\
$^1$Aryabhatta Research Institute of Observational Sciences, Manora Peak, Nainital - 263 129\\
$^2$Raman Research Institute, Bangalore - 560 080 \\
$^3$Dept. of Physics, Indian Institute of Science, Bangalore}
\date{Received ---; accepted ---}
\maketitle
\label{firstpage}
\begin{abstract}
We present the CCD Cousins R band photometric observations of the afterglow of GRB 041006. The multiband afterglow evolution is modelled using an underlying `hard' electron energy spectrum with a $p_1 \sim 1.3$. The burst appears to be of very low energy ($E \sim 10^{48}$~ergs) confined to a narrow cone of opening angle $\theta \sim 2.3^{\circ}$. The associated supernova is compared with SN1998bw and is found to be brighter. 
\end{abstract}
\begin{keywords}
Gamma-ray bursts, GRB afterglows, flux decay, spectral index
\end{keywords}
%
%
\section{Introduction}
GRB 041006 was detected by the High Energy Transient Explorer (HETE II) Fregate and WXM 
instruments at 12:18:08 UT on 06 October 2004 (Galassi et al. 2004).
The refined error circle provided by the HETE II ground analysis was 
of radius 5 arc minutes centered at 
$\alpha$ = $00\rm^{h} 54\rm^{m} 53\rm^{s}$, $\delta$ = $+01^{\circ} 12^{\arcmin} 04^{\arcsec}$
 (J2000). The fluence of GRB 041006 in 2$-$30 keV and 30$-$400 keV were 5 $\times$ 10$^{-6}$
 erg/cm$^{2}$ and 7 $\times$ 10$^{-6}$ erg/cm$^{2}$ respectively which classifies this as an 
``x-ray rich GRB". The light curve shape and the spectral characteristics of GRB 041006
 were similar to GRB 030329 though it was 20 times fainter than GRB 030329. 
A fading x-ray counterpart was reported by Butler et al. (2004) and the optical afterglow 
located at  
$\alpha$ = $00\rm^{h} 54\rm^{m} 50\rm^{s}.17$, $\delta$ = $+01^{\circ} 14^{\arcmin} 07^{\arcsec}$
 (J2000) was found by Da Costa, Noel and Price (2004). 
From low resolution spectroscopy performed on October 7.10 UT
Fugazza et al. (2004) derived a redshift of 
z = 0.712 for this burst. This value was later confirmed by Price et al. (2004). 
The optical afterglow showed a `supernova bump'  around $5.0$~ days. Adopting the fluence reported by Galassi et al. (2004) and the estimated redshift, Fugazza et al. (2004) derive the isotropic equivalent gamma$-$ray energy of the burst to be $E_{iso}$ = 1.5 $\times$ 10$^{52}$ erg. 

In this paper we present Cousins R filter optical observations of the afterglow of 
GRB 041006 carried out during the first night of the GRB trigger and the modelling of the data along with the multiband observations available so far in the literature. For a reliable determination of the 
optical afterglow magnitudes we imaged the field of GRB 041006 along with RU 152 
standard region of Landolt (1992) and calibrated a total of 10 stars in the field. 
The optical observations and the data reduction is discussed in section 2 and the modelling assuming a `double slope' electron energy spectrum is discussed in section 3. Section 4 explains briefly about the associated supernova and its comparison with SN1998bw. A brief discussion is presented in section 5 and the conclusions follow in section 6.
%
%
\section{Observations and Data Reduction} 
The observations in Cousins R filter of the optical afterglow of GRB 041006 were 
carried out on 06 October 2004 using the 104-cm Sampurnanand Telescope of ARIES, 
Nainital. One pixel of the 2048 $\times$ 2048 pixel$^2$ size chip corresponds to a 
square of $0\arcsec.38$ side, and the entire chip covers a field of 13$\arcmin$ $\times$ 
13$\arcmin$ on the sky. The gain and read out noise of the CCD camera are 10 e$^-$/ADU
 and 5.3 e$^-$ respectively.

We observed the Landolt (1992) standard RU 152 and the OA field in VRI filters on 
08 October 2004 for photometric calibration during good photometric sky conditions. 
Several bias and twilight flat frames were observed with the CCD camera to calibrate the 
images using standard techniques.

ESO MIDAS, NOAO IRAF and DAOPHOT softwares were used to process the images. The values 
of the atmospheric extinction on the night of 08/09 October 2004 determined from the 
observations of RU 152 bright stars are 0.19, 0.15 and 0.11 mag in V, R and I
filters respectively. 
The 7 standard stars in the RU 152 region cover a range of -0.06 $< (V-R) <$ 0.47
in color and of 11.08 $< V <$ 15.02 in 
brightness. This gives the following transformation equations\\

\noindent 
$v_{ccd}$ = $V - (0.11 \pm 0.01) (V-R) + (4.29 \pm 0.004)$\\
$r_{ccd}$ = $R - (0.12 \pm 0.01) (V-R) + (4.20 \pm 0.004)$\\
$i_{ccd}$ = $I - (0.14 \pm 0.02) (R-I) + (4.69 \pm 0.007)$\\

\noindent 
where V, R, I are standard magnitudes and $v_{ccd}$, $r_{ccd}$, $i_{ccd}$ represent the
instrumental magnitudes normalized for 1 s of exposure time and corrected for 
atmospheric extinction. The color coefficients, zero-points and errors in them were 
determined by fitting least square linear regressions to the data points. Using these 
transformations, VRI photometric magnitudes of 10 secondary stars were determined in 
GRB 041006 field and their values are listed in Table 1. The ($X, Y$) CCD pixel 
coordinates are converted to $\alpha_{2000}$, $\delta_{2000}$ values using the astrometric
positions given by Henden (2004). The 10 secondary stars in the GRB 041006 field were 
observed 2, 4 and 2 times in $V, R$ and $I$ filters respectively. These stars have internal 
photometric 
accuracy better than 0.01 mag. The zero-point differences on comparison between our
photometry and that of Henden's (2004) are 0.058 $\pm$ 0.043, 0.062 $\pm$ 0.045 and 
0.026 $\pm$ 0.041 in $V, R$ and $I$ filters respectively. These differences are based 
on the comparison of the 10 secondary stars in the GRB 041006 field.

Our first observation was made at $\sim$ 0.24 days after the burst. Several short exposures, with exposure 
time varying from 5 min to 25 min, were taken to image the OA of GRB 041006. To improve the 
signal-to-noise ratio of the OA, the data have been binned in 2$\times$2 pixel$^{2}$. 
Profile-fitting magnitudes were determined from these images using DAOPHOT II software. 
The profile magnitudes were converted to aperture (about 4 arcsec) magnitudes using aperture
growth curve determined from well isolated secondary stars. They were differentially calibrated 
with respect to the secondary stars listed in Table 1. The magnitudes derived in this way are
given in Table 2. 

The published photometric measurements by Kinugasa et al. (2004a), Ferrero et al. (2004), Monfardini et al. (2004), Fynbo et al. (2004), Fugazza et al. (2004), Greco et al. (2004), D'Avanzo et al. (2004), Garnavich et al. (2004), Balman et al. (2004), Covino et al. (2004), Bikmaev et al. (2004) and
Garg et al. (2004) were converted to the present photometric scales using the secondary stars listed in Table 1
%
%
\begin{table}[h]
\caption{The identification number (ID), ($\alpha$, $\delta$) for epoch 2000, 
standard $V, (V-R)$ and $(R-I)$ photometric magnitudes of the stars in the 
GRB 041006 region are given. Number of observations taken in $V, R$ and $I$ 
filters are 2, 4 and 2 respectively.}
\medskip
\begin{center}
\begin{tabular}{cccccc} \hline
ID&  $\alpha_{2000}$& $\delta_{2000}$ & $V$ & $V-R$ & $R-I$\\
&(h m s)& (deg m s)& (mag)& (mag)& (mag)\\ \hline
1& 00 54 47&01 14 05&17.99&0.19&0.27\\
2& 00 54 37&01 14 02&13.67&0.33&0.39\\
3& 00 55 01&01 11 25&13.98&0.43&0.40\\
4& 00 54 38&01 12 59&17.36&0.57&0.54\\
5& 00 54 54&01 15 47&16.96&0.59&0.56\\
6& 00 54 53&01 17 47&15.19&0.58&0.53\\
7& 00 54 55&01 16 58&15.21&0.42&0.40\\
8& 00 54 57&01 19 31&14.38&0.60&0.53\\
9& 00 55 05&01 17 07&16.21&0.38&0.36\\
10&00 55 01&01 15 39&15.47&0.47&0.47\\ \hline
\end{tabular} 
\end{center}
\end{table}
%
%
\begin{table}[h]
\caption{CCD Cousins R band optical observations of the GRB 041006 afterglow using the 104-cm 
Sampurnanand Telescope at ARIES, Nainital.}
\medskip
\begin{center}
\begin{tabular}{cccc} \hline
Date (UT)& Magnitude & Exposure Time  & Passband \\
2004 October&(mag)& (s)& \\ \hline
6.7572 & 19.80 $\pm$ 0.056 &300&R\\
6.7666 & 19.74 $\pm$ 0.034 &600&R\\
6.7763 & 19.91 $\pm$ 0.043 &600&R\\
6.7867 & 19.93 $\pm$ 0.053 &800&R\\
6.8151 & 20.02 $\pm$ 0.047 &1200&R\\
6.8305 & 20.06 $\pm$ 0.059 &1200&R\\
6.8472 & 20.04 $\pm$ 0.047 &1200&R\\
6.8624 & 20.06 $\pm$ 0.059 &1200&R\\
6.8772 & 20.23 $\pm$ 0.062 &1200&R\\
6.9101 & 20.22 $\pm$ 0.053 &1500&R\\ \hline
\end{tabular} 
\end{center}
\end{table}
%
%
\begin{figure}
\hspace{2.0cm}
\psfig{file=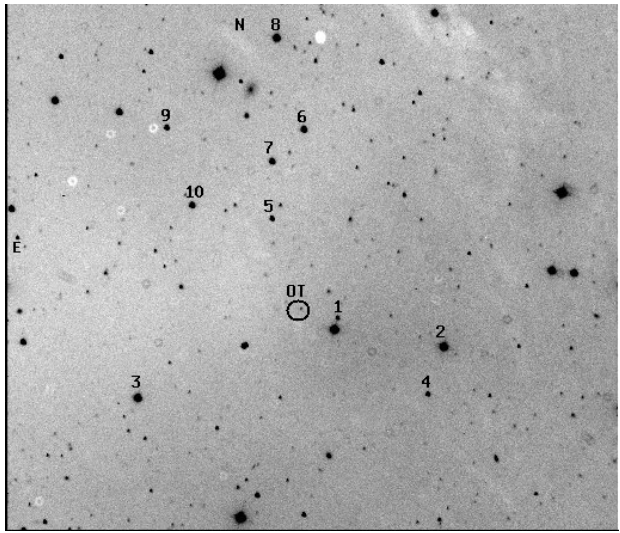,height=7.0cm}
\vspace{0.8cm}
\caption{A 12$\arcmin$ $\times$ 12$\arcmin$ R band image of GRB 041006 optical afterglow. The position 
of the optical afterglow is marked with a circle. The 10 secondary stars used for calibration are also 
marked. North is up and East is to the left.}
\end{figure}
%
%
\section{Multiband Modelling}
The afterglow flux $f_{\nu}(t)$ observed at a given frequency $\nu$ at time $t$ after the burst is usually described by a powerlaw ($f_{\nu}(t) \propto \nu^{-\beta} t^{-\alpha}$), with $\alpha$ and $\beta$ assuming different values in different spectral and dynamical regimes of evolution (See Piran 2004 for a recent review).
 The optical afterglow of GRB041006 exhibited a relatively flat decay $\alpha_1 \sim 0.5$ in the early phase which steepened to $\alpha_2 \sim 1.3$ after a break around $0.14$~day (Stanek et al. 2004, Soderberg et al. 2005). If this break is due to the lateral spreading of a conical fireball, the indices indicate a relatively hard ($p < 2$) energy spectrum for the synchrotron emitting electrons (Bhattacharya 2001). At $0.058$~day the intrinsic spectral index of the afterglow was $0.685 \pm 0.13$ between $B$ and $V$ bands (Da Costa \& Noel 2004). At $1.74$ day, the index became $1.0 \pm 0.18$ between $B$ and $R$ bands (Garnavich et al. 2004). But this is only a marginal steepening, seen within one sigma error. The x-ray spectral index of $0.9 \pm 0.2$ measured at $1.23$ days (Butler et al. 2004) is consistent with the $\beta$ observed within the optical bands. Hence, the reported spectral indices in both optical and x-ray bands support the inference of a hard electron energy spectrum.  Butler et al. (2004) report that the x-ray lightcurve follows a single power law of slope $-1.0 \pm 0.1$ during $0.7$ to $1.7$~days. This indicates that the x-ray flux decay is slower than the optical. But using the same data presented by the authors we find that a powerlaw of slope even close to $1.4$ could describe the flux evolution fairly well. Apart from the optical and x-ray observations, upper limits to the afterglow flux are available in near IR (Kinugasa et al. 2004b), millimeter (Barnard et al. 2004a, 2004b) and also in radio (Soderberg et al. 2004) wavelengths.
 
 We attempted modelling the multi wavelength afterglow evolution with an underlying `double slope' electron energy spectrum. The electrons are assumed to be distributed in a flatter spectrum ($p_{1} < 2$) until a cutoff energy $\gamma_i m_e c^2$, and thereafter towards the higher energy end, the distribution steepens to $p_{2} >2$ (see Panaitescu and Kumar (2001) for a discussion on such double slope electron energy distribution). $\gamma_i$ is termed as the `injection break'. The evolution of $\gamma_i$ significantly affects the afterglow light curve. Following Bhattacharya (2001), we parametrize the evolution of the injection break as $\gamma_i$ = $\xi \Gamma^q$ where $\Gamma$ is the bulk Lorentz factor of the shock and $\xi$ is the constant of proportionality. 

 We assume the cooling frequency ($\nu_c$) to be below the optical bands to satisfy  $\alpha$ of $0.5$ and $\beta$ in the range of $0.6 - 0.7$ simultaneously. There is no signature of a steepening seen at the higher energy end of the spectrum from the available observations. The upper cut off of the electron energy spectrum may lie at energies higher than what the observations could probe. Hence we place the `injection break frequency', $\nu_i$, corresponding to the electron Lorentz factor $\gamma_{i}$, above the x-ray band. We compute the spectral evolution of the afterglow with these basic assumptions. A broken power-law model for the underlying spectrum and evolution is used (Sari et al. 1998 ), with Band-type smoothening (Band et al. 1993) applied to spectral and temporal breaks. After excluding the NIR/mm/radio upper-limits and optical data beyond $5$ days (where a supernova bump prominently appears) from the fit, the model converged to a minimum $\chi^2_{\rm{DOF}}$ of $3.4$. An additional extinction other than that due to Galactic dust ($A_V$ of $0.07$~mag, estimated for this direction using the data of Schlegel et al. 1989) is required by the model, which could result from the dust distribution in the host galaxy of the GRB. We obtained  $A_V$ in the range of $0.03 - 0.16$~mag in the host rest frame, assuming a Galaxy type extinction law for the dust distribution (Cardelli et al. 1989). According to Butler et al. (2004) the x-ray spectrum is well fit by an absorbed power-law with the neutral hydrogen column density ($N_H$) greater than that expected from the Galactic column along the line of sight. They estimate an excess $N_H$ of $3.2 \times 10^{21}$~${\rm{cm}}^{-2}$ for the host rest frame. The rest frame $N_{H}$ to $E_{(B-V)}$ ratio is two to three orders of magnitude higher than the empirical value obtained for the ISM of our galaxy (Predehl \& Schmitt, 1994). This might imply a different type of dust in the GRB host, and is not a surprising fact if GRBs are to occur at star forming regions, where dust depletion is highly probable. 

In determining the physical parameters, the total number of unknowns were five (isotropic equivalent energy release $E_{\rm{iso}}$, number density of the ambient medium $n$, fractional energy content in electrons $\epsilon_e$, that in the magnetic field $\epsilon_B$ and $\xi$, the constant of proportionality for $\gamma_i$. However the total number of parameters inferred from the observations are four since the self absorption frequency $\nu_a$ is not well determined. We hence assumed a range of values for $n$, and obtained the remaining four physical parameters for the afterglow. Since only lower limit is available for $\nu_i$ and $p_2$, $\xi$ is not well constrained from the upper end. This will also affect the precise determination of $\epsilon_e$, but the best fit provides a value close to unity for $\epsilon_e$. Table 3 \& 4 lists the fit parameters and the derived physical parameters for a nominal value of $n = 1.0$ respectively, from the best fit model. A decrease in the value of $n$ to 0.05 atom/cc increases $E_{iso}$ to $8 \times 10^{52}$~ergs, and $\epsilon_B$ to $0.2$. By increasing the number density to $50$~atom/cc, the isotropic energy decreased to $10^{51}$~ergs and $\epsilon_B$ decreased to $3. \times 10^{-3}$. We obtain a range of $(1.0 - 7.0) \times 10^{48}$~ergs for the total energy release corresponding to a range of $0.05 - 50.0$~atom/cc for $n$. We used z = 0.716 (Price et al. 2004) and a corresponding luminosity distance of 4.72 Gpc (assuming a cosmology of H$_{0}$ = 65 km/sec/Mpc, $\Omega_{m}$ = 0.3 and $\Omega_{\Lambda}$ = 0.7). We have displayed the observations along with the model in figure 2.
%
%
\begin{table}
\caption{Model Parameters}
\begin{center}
\begin{tabular}{|c|c||c|c|} \hline
\multicolumn{2}{|c|}{Electron Distribution}&
    \multicolumn{2}{|c|}{Spectral} \\ 
\multicolumn{2}{|c|}{Parameters}&
    \multicolumn{2}{|c|}{Parameters} \\ 
\multicolumn{2}{|c|}{}&
    \multicolumn{2}{|c|}{(At $0.1$ days)} \\ \hline \hline
$p_1$ & $1.29- 1.32$ & $\nu_m$&$ (1.2 - 3.0) \times 10^{12}$~Hz \\ \hline
$p_2$& $> 2.2$&$\nu_c$&$ (1.0 - 2.0) \times 10^{14}$~Hz \\ \hline
$q$& $1.0$&$\nu_i$&$ <2.4 \times 10^{20}$~Hz \\ \hline
$t_j$&$0.17 - 0.24$~day&$f_m$&$(0.37 - 0.49)$~mJy \\ \hline
\multicolumn{4}{|c|}{Others}\\ \hline
$E_{(B-V)}$~(host)&$0.01 - 0.05$~mag &&\\ \hline
\end{tabular}
\end{center}
\end{table}
%
%
\begin{table}
\caption{Physical Parameters}
\begin{center}
\begin{tabular}{|c|c|} \hline
$n$(assumed)&$1.0$~atom/cc \\ \hline
$\epsilon_{e}$&$\sim 1.0 $\\ \hline
$\epsilon_{B}$&$0.07 - 0.14$\\ \hline
$\xi$&$ > 2.0 \times 10^{5}$ \\ \hline
$E_{iso}$&$(2.0-4.0) \times 10^{51}$~ergs \\ \hline
$\theta_j$&$1.7^{\circ} - 2.8^{\circ}$ \\ \hline
$E_{tot}$&$(1.4 -3.4) \times 10^{48}$~ergs\\ \hline
\end{tabular}
\end{center}
\end{table}
\begin{figure}
\begin{center}
\psfig{file=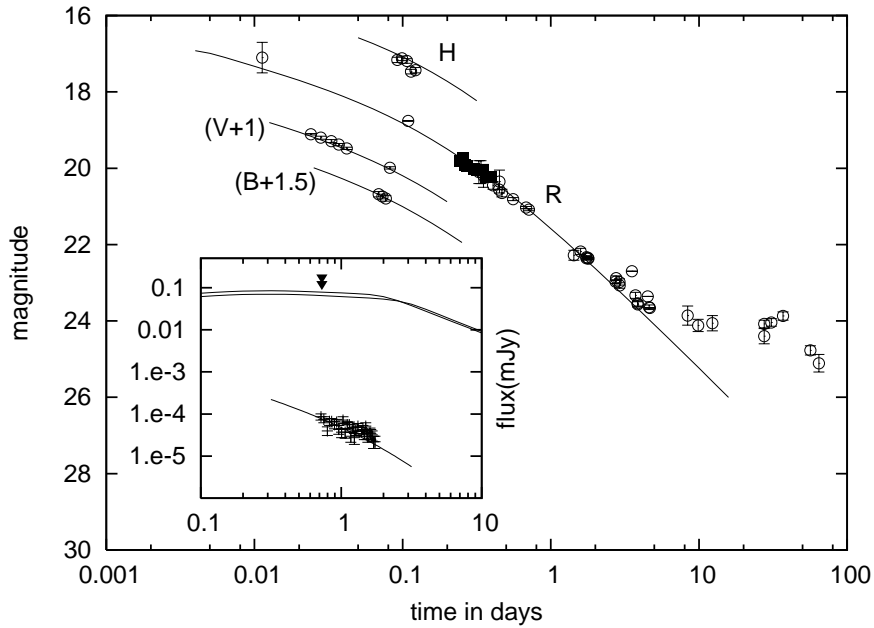, height=8.5cm}
\end{center}
\caption{GRB 041006 afterglow observations in different bands along with the model. Filled circles represent our data and open circles represent the data from literature. Light curve in different bands are shifted as indicated. Inset : X-ray observations and radio upper limits along with the model.}
\end{figure}

 Since available observations do not cover the 0.12 day break in various bands, its achromatic nature as expected for a jet break, cannot be confirmed. As a mild evolution in the optical spectral index between $0.06$ days to $1.7$ days has also been reported, we examined the possibility of the temporal steepening arising from the passage of a spectral break through R-band. The observed change in $\alpha$ is $\sim 0.8$ and the maximum change that could be attributed to $\beta$ is $0.3$. Passage of the cooling frequency would predict $\Delta \alpha = 0.25$ and $\Delta \beta = 0.5$, and hence can not reproduce the observed $\alpha$ and $\beta$. The only other possibility is the passage of the injection break through R-band around $0.1$ day, with $p_1$ being $\sim 1.3$ and $p_2$ $\sim 2.3$. A model with this assumption can reproduce the optical observations and the NIR/mm/radio upper limits very well but it underpredicts the x-ray observations by two orders of magnitude, requiring a substantial contribution from inverse-compton emission. Following the method used by Sari \& Esin (2001), appropriately modified for $p < 2$, we calculated the inverse compton emission for this model, and found that it is negligible at the x-ray frequencies.   

 We find from our analysis that the best possible explanation for the multi wavelength observations for this afterglow is the synchrotron emission from a collimated outflow with electrons distributed in a hard energy spectrum. 
%
%
\section{The associated supernova}
The afterglow optical flux decay has shown deviation from a powerlaw behaviour after $ \sim 0.5$ days and exhibited a bump around 25 days. Stanek et al. (2004) attributes this behaviour to the emission from an associated supernova. We subtracted the modelled power-law flux and an assumed host galaxy flux from the observed emission to estimate the supernova contribution. There is no measurement available for the host galaxy flux and we are left with only assumptions. Stanek et al. (2004) find that the supernova associated with GRB 041006 peaks at a later (1.35 times) time after explosion and is also
brighter in comparison to SN1998bw redshifted to $z$ $=$ $0.716$. Soderberg et al. (2005) from the late HST observations (mainly I band) conclude that the supernova is $\sim 0.3$ magnitudes brighter than a k-corrected SN1998bw whilst no additional stretch in time is required. They also point out the possibility of late time flux getting contaminated by the host and also by a nearby galaxy.

The major difference our afterglow model has, in this context, is the presence of the host extinction (rest frame $A_V=0.1$~mag in the best fit model). 
We used the late afterglow magnitudes of Stanek et al. (2004) in R-band and Soderberg et al. (2005) in I band to obtain the residual flux, which could be attributed to the associated supernova after subtracting the host flux. 
We assumed four free parameters (host galaxy flux in both the bands, $\Delta m$, the
  shift in magnitude required and $\delta$, the multiplication factor for
  the time axis) and performed a linear least square fit. The template
  matches  best with the associated SN lightcurve for $\Delta m$ in the
  range (one sigma) of $-0.55$ to $-0.75$, $\delta$ between $1.05$ and $1.1$, R(host) in the
  range of $26.2$ to $28.8$~mag and I(host) between $26.8$ and $28.9$~mag (figure 3).
From the figure, it appears that the I band data fits better with the
  template compared to the R band.
  We notice that a different value of $\Delta m$ and $\delta$ would be
  required to fit the R band better, which suggests that the associated SN
  may have a different spectrum compared to that of SN1998bw.
This could also be the reason for the different conclusions arrived at by Stanek et al. (2004) and Soderberg et al. (2005) about the associated SN.
%
%
\begin{figure}
\begin{center}
\psfig{file=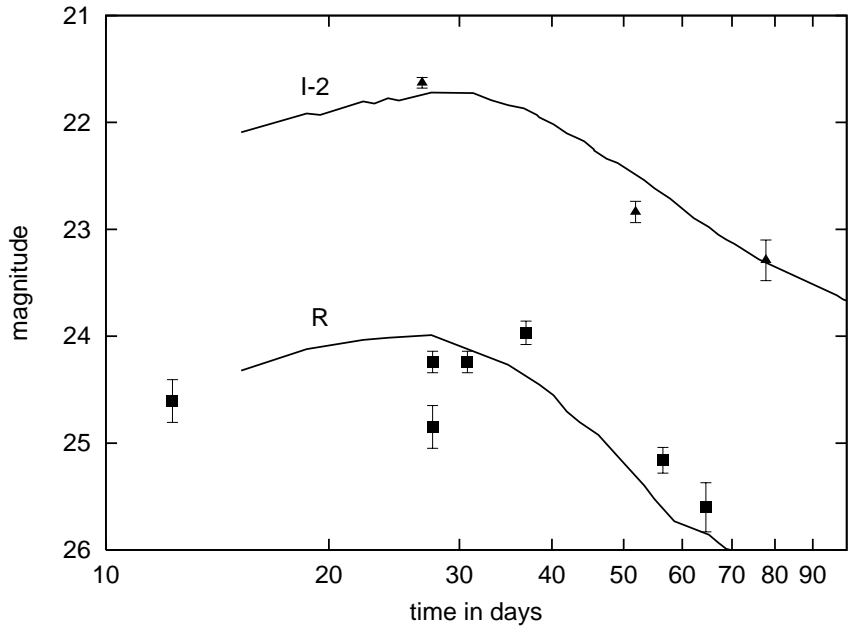,height=8.5cm}
\end{center}
\caption{Comparison between the associated supernova in R (squares) and I (triangles) bands with the redshifted k-corrected SN1998bw (line) after applying a $\Delta m$ and $\delta$ correction (see text). The parameters used in obtaining the data sets are $\Delta m = -0.65$, $\delta = 1.05$, R(host) $= 26.2$ and I(host) $= 28.9$. The curves have been shifted as indicated.} 
\end{figure} 
%
%
\section{Discussion}
The decay index of $\sim 1.3$ seen in GRB 041006 afterglow lightcurve, post the one day break, suggests that the underlying electron energy spectrum could be hard ($p \sim 1.3$). A different model,
proposed by Granot, Ramirez-Ruiz \& Perna (2005), attributes this behaviour to the jet being viewed off-axis. In this model, the $ \sim 0.1$~day break is caused by the lorentz factor falling below the inverse of the viewing angle, there is no lateral expansion of the jet and the derived electron 
energy distribution is steeper $(p \sim 2.2)$. They invoke the presence of a stellar wind driven ambient density profile to explain the reported x-ray decay index ($\alpha_x \sim 1.1$) which is flatter in comparison to the simultaneous optical decay index ($\alpha_o \sim 1.3$).
Granot et al. (2005) derive a kinetic energy of $10^{51}$~ergs for this explosion. 

Since we do not find the x-ray decay index to be flat (see section 3.), we used a normal ISM density profile for the ambient medium. The kinetic energy derived for the explosion is rather small ($10^{48}$~ergs). 
 With the available set of data, it is impossible to discriminate between the two models.   
However, both off-axis jets and wind driven density profiles, although realistic, are not inferred commonly from the afterglow models. For this afterglow, in fact, there is no need of going beyond the simplistic assumption of an on-axis jet ploughing through a constant density interstellar medium, if one assumes a flatter electron energy spectrum with an appropriate upper cut off at higher energies. 

Butler et al. (2004) extrapolate the observed optical spectrum, which has an index $\beta_{\rm{o}}$ of $-1.0 \pm 0.1$ to the x-ray frequencies and find that the observed x-ray flux is underpredicted. They suggest that a possible explanation for this and also for the smaller $\alpha_x$ could be the presence of considerable amount of inverse compton emission in the x-ray bands. They also find that the spectral index within the x-ray band ($\beta_x \sim 0.7$) is smaller than that within the optical band ($\beta_o \approx -1.1 \pm 0.1$). Since in our model, we have a hard electron energy spectrum, $\beta$ is close to $0.7$, and the presence of extinction in optical bands due to the dust column of the host galaxy, explains the observed higher optical $\beta$. Hence our model does not underpredict the observe x-ray flux, eventhough, the inverse compton emission predicted by the model is negligible at the x-ray frequencies.

%
\section{Conclusions}
\begin{itemize}
\item In this paper, we present the $R_c$ band optical observations of the afterglow. 
\item We also show that multi band observations can be well reproduced by synchrotron emission from a hard electron energy spectrum and an early jet break ($t_j = 0.2$~day). 
\item The model requires additional extinction due to the dust distribution in the host. The E(B-V) derived by us is two orders of magnitude lesser than what one would predict from the $N_H$ column density inferred by Butler et al. (2004) assuming a Galaxy type gas to dust ratio. This difference might indicate a considerable difference in the characteristics of dust column of the host.
\item The kinetic energy of the burst inferred from the model is low ($\sim 10^{48}$~ergs), and is an order of magnitude smaller than the energy emitted in gamma rays. 
\item Comparison of the afterglow subtracted optical flux beyond $5.0$~days with a redshifted k-corrected SN1998bw light curve shows that the associated supernova is brighter than SN1998bw, and has a longer rise time and might have a different spectrum than SN1998bw.
\end{itemize}
%
%
\section*{Acknowledgments}
This research has made use of data obtained through the High Energy Astrophysics Science
Archive Research Center Online Service, provided by the NASA/Goddard Space Flight Center. 
The GCN system, managed and operated by Scott Barthelmy, is gratefully acknowledged. 
LR acknowledges Council for Scientific and Industrial Research, India for financial support.
We are thankful to an anonymous referee for useful comments which improved the manuscript.

\clearpage
\newpage

\label{lastpage}
\end{document}